\documentclass[preprint,12pt]{elsarticle}
\usepackage{amssymb}
\usepackage{graphicx}% Include figure files
\journal{Physics Letters B}

\begin{document}

\begin{frontmatter}



\title{Probing the $f(R)$ formalism through gravitational wave polarizations}


\author{M. E. S. Alves, O. D. Miranda and J. C. N. de Araujo}

\address{INPE-Instituto Nacional de Pesquisas Espaciais - Divis\~ao
de Astrof\'isica,
\\Av.dos Astronautas 1758, S\~ao Jos\'e dos Campos, 12227-010 SP, Brazil\\}
\ead{alvesmes@das.inpe.br, oswaldo@das.inpe.br,
jcarlos@das.inpe.br}

\begin{abstract}
The direct observation of gravitational waves (GW) in the near
future, and the corresponding determination of the number of
independent polarizations, is a powerful tool to test general
relativity and alternative theories of gravity. In the present
work we use the Newman-Penrose formalism to characterize GWs in
quadratic gravity and in a particular class of $f(R)$ Lagrangians.
We find that both quadratic gravity and the $f(R)$ theory belong
to the most general invariant class of GWs, i.e., they can present
up to six independent polarizations of GWs. For a particular
combination of the parameters, we find that quadratic gravity can
present up to five polarizations states. On the other hand, if we
use the Palatini approach for $f(R)$ theories, GWs present only
the usual two transverse-traceless polarizations such as in
general relativity. Thus, we conclude that the observation of GWs
can strongly constrain the suitable formalism for these theories.
\end{abstract}


\end{frontmatter}

\section{Introduction}

Modifications of Einstein's gravity have been considered in
several approaches. The Lagrangians which consider higher orders
of the Ricci scalar $R$ and the Ricci tensor $R_{\mu\nu}$ have
been proposed as extensions of general relativity \cite{Weyl1922}.
The semiclassical theory considers the backreaction of quantum
fields in a classical geometric background
\cite{deWitt1965,Hu2004}. Such Lagrangians predict field equations
with four orders derivatives of the metric, rather than the two
orders derivatives in the general relativity theory
\cite{deWitt1965,Birrel1982}. Quadratic Lagrangians have been used
also to yield renormalizable theories of gravity coupled to matter
\cite{stelle1977}. Moreover, higher-derivative theories arise as a
low energy limit of string theories
\cite{Zwiebach1985,Tseytlin1986}.

Starobinsky \cite{Starobinsky1980} argues that the higher order
terms could mimic a cosmological constant. Recently, this idea has
been largely studied as a potential way to address the dark energy
problem. In this context, several different forms of the modified
Lagrangians have been recently considered constituting a class of
theories, the so-called $f(R)$ theories (see e.g. \cite{several}
and references therein).

In the context of gravitational waves (GWs), it was shown that
quadratic gravity\footnote{For the sake of definiteness we will
call quadratic gravity that theory which takes into account not
only $R^2$ but also $R_{\mu\nu}R^{\mu\nu}$ in the Lagrangian.}
presents a frequency-dependent shift in the wave amplitude with
respect to the wave amplitude obtained from general relativity
\cite{deReyNeto2003}. For the $f(R)$ theories, it was found that
GWs can have a massive-like scalar mode besides the usual
transverse-traceless modes \cite{Capozziello}.

A powerful tool to study the properties of GWs in any metric
theory of gravity was developed by Eardley \emph{et al.}
\cite{Eardley1973}. The basic idea is to analyze all the
physically relevant components of the Riemann tensor
$R_{\lambda\mu\kappa\nu}$, which cause relative acceleration
between test particles. The GWs in a metric theory involves the
metric field $g_{\mu\nu}$ and any auxiliary gravitational fields
that could exist. But the resultant Riemann tensor is the only
measurable field.

In their work, Eardley \emph{et al.} used a null-tetrad basis in
order to calculate the Newman-Penrose \cite{Newman} quantities in
terms of the irreducible parts of $R_{\lambda\mu\kappa\nu}$,
namely, the Weyl tensor, the tracelless Ricci tensor and the Ricci
scalar. This analysis showed that there are six possible modes of
polarization of GWs in the most general case, which can be
completely resolved by feasible experiments. Thus, it is possible
to classify a given theory by the non-null Newman-Penrose
quantities \cite{Eardley1973}.

The aim of the present work is to characterize GWs for a
particular class of $f(R)$ gravity and for quadratic gravity
making use of the Newman-Penrose formalism. Since the field
equations derived from such Lagrangians in the metric formalism
yield dynamical equations for $R$ and $R_{\mu\nu}$, the analysis
consists basically to find the resultant expressions for $R$ and
$R_{\mu\nu}$ in the week field limit, without the need to write
explicitly the expressions in terms of the metric perturbations.
This makes the classification easier and clear. We also mention
how the use of the Palatini approach in deriving the field
equations for $f(R)$ gravity affect the number of independent
polarizations of GWs. This comes from the fact that the
classification of a given $f(R)$ is formalism-dependent.

It is argued that the observations of the GWs in the near future
(for the current status of GWs detectors see e.g.
\cite{virgo2007,ligo,ligo2005,tama2002,aguiar2008}), and the
corresponding determination of all possible states of
polarization, is a very powerful test for the present studied
alternative theories of gravity. Particularly, we show that GWs
experiments can be decisive for quadratic gravity and in the
determination of the suitable formalism for $f(R)$ theories, i.e.,
the use of the metric or the Palatini approaches, since the number
of polarizations of GWs depends on the formalism
used\footnote{See, in particular, \cite{dePaula2005} for an
application of the Newman-Penrose formalism to determine the GW
polarizations in massive gravity.}.

\section{The Newman-Penrose formalism - an overview}\label{revisit}

\begin{figure}[!ht]
\centering
\includegraphics[width=90mm]{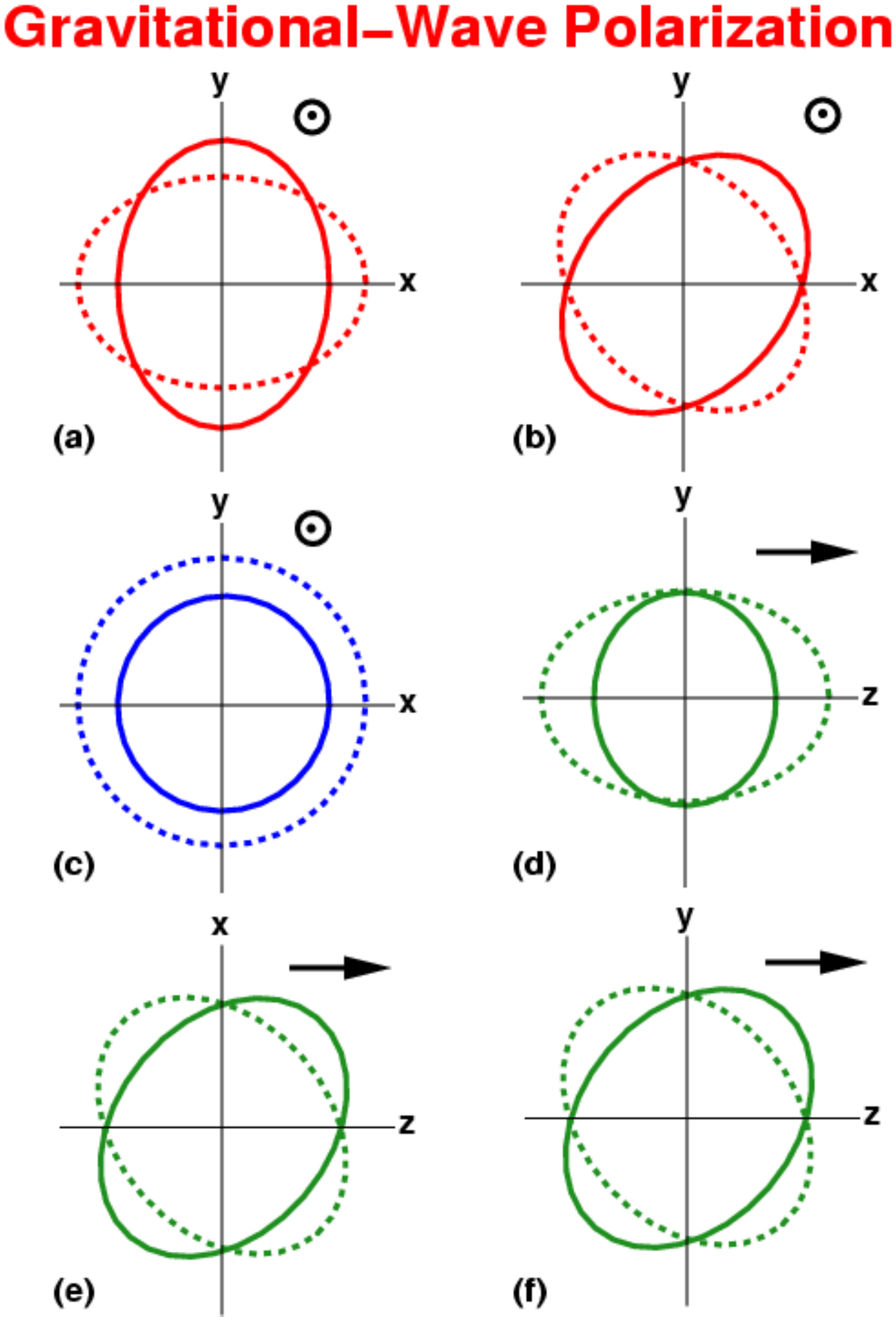}
\caption{The six polarization modes of weak, plane,
null GW permitted in any metric theory of gravity.
Also shown is the displacement that each mode induces on a sphere of test particles. The wave
propagates out of the plane in $(a)$, $(b)$ and $(c)$, and it propagates in the plane
in $(d)$, $(e)$ and $(f)$. The displacement induced on the sphere of test particles
corresponds to the following Newman–Penrose quantities: Re$\Psi_4(a)$, Im$\Psi_4(b)$,
$\Phi_{22}(c)$, $\Psi_2(d)$, Re$\Psi_3(e)$, Im$\Psi_3(f)$.}
\label{fig1}
\end{figure}

Throughout this paper we consider GWs propagating in the $+z$
direction. So, all the quantities are functions only of $t$ and
$z$.
\\
At any point $P$, the null complex tetrad $(\textbf{k},
\textbf{l}, \textbf{m}, \overline{\textbf{m}})$ is related to the
Cartesian tetrad $(\textbf{e}_t, \textbf{e}_x, \textbf{e}_y,
\textbf{e}_z)$ by:
\begin{equation}
\textbf{k} = \frac{1}{\sqrt{2}}( \textbf{e}_t + \textbf{e}_z )
\end{equation}
\begin{equation}
\textbf{l} = \frac{1}{\sqrt{2}}( \textbf{e}_t - \textbf{e}_z )
\end{equation}
\begin{equation}
\textbf{m} = \frac{1}{\sqrt{2}}( \textbf{e}_x + i \textbf{e}_y )
\end{equation}
\begin{equation}
\overline{\textbf{m}} = \frac{1}{\sqrt{2}}( \textbf{e}_x - i\textbf{e}_y )
\end{equation}
It is easy to verify that the tetrad vectors obey the relations:
\begin{equation}
-\textbf{k} \cdot \textbf{l} = \textbf{m} \cdot \overline{\textbf{m}} = 1
\end{equation}
\begin{equation}
\textbf{k} \cdot \textbf{m} = \textbf{k} \cdot \overline{\textbf{m}} =
\textbf{l} \cdot \textbf{m} = \textbf{l} \cdot \overline{\textbf{m}} = 0
\end{equation}
The null-tetrad components of a tensor \textbf{T} are written
according to the notation:
\begin{equation}
T_{abc \cdot \cdot \cdot } = T_{\mu\nu\lambda \cdot \cdot \cdot }a^\mu b^\nu c^\lambda \cdot\cdot\cdot,
\end{equation}
where $(a,b,c,\cdot \cdot \cdot)$ run over $(\textbf{k},
\textbf{l}, \textbf{m}, \overline{\textbf{m}})$ and
$(\mu,\nu,\lambda,\cdot \cdot \cdot)$ run over $(t,x,y,z)$ since
we are working in cartesian coordinates.
\\
In general the Newman-Penrose quantities, namely the ten $\Psi$'s,
nine $\Phi$'s, and $\Lambda$, which represent the irreducible
parts of the Riemann tensor $R_{\lambda\mu\kappa\nu}$, are all
algebraically independent. When we restrict ourselves to nearly
plane waves, however, we find that the differential and symmetry
properties of $R_{\lambda\mu\kappa\nu}$ reduce the number of
independent, nonvanishing components, to six. Thus, we shall
choose the set $\{\Psi_2,\Psi_3,\Psi_4,\Phi_{22}\}$ to describe,
in a given null frame, the six independent components of a wave in
the generic metric theory. In the tetrad basis, the Newman-Penrose
quantities of the Riemann tensor are, therefore, given by:
\begin{equation}
\Psi_2 = -\frac{1}{6} R_{lklk},
\end{equation}
\begin{equation}
\Psi_3 = -\frac{1}{2} R_{lkl\overline{m}},
\end{equation}
\begin{equation}
\Psi_4 = - R_{l\overline{m}l\overline{m}},
\end{equation}
\begin{equation}
\Phi_{22} = - R_{lml\overline{m}}.
\end{equation}
Note that, $\Psi_3$ and $\Psi_4$ are complex, thus each one
represents two independent polarizations. One polarization for the
real part and one for the imaginary part, totalizing six
components (see fig. \ref{fig1} which was taken from
\cite{Will2006}).

Analyzing the behaviour of the set
$\{\Psi_2,\Psi_3,\Psi_4,\Phi_{22}\}$ under rotations, we see that
they have the respective helicity values $s=\{0, \pm 1, \pm 2, 0
\}$.

We also have the very useful relations for the Ricci tensor:
\begin{equation}
R_{lk} = R_{lklk},
\end{equation}
\begin{equation}
R_{ll} = 2 R_{lml\overline{m}},
\end{equation}
\begin{equation}
R_{lm} = R_{lklm},
\end{equation}
\begin{equation}
R_{l\overline{m}} = R_{lkl\overline{m}},
\end{equation}
and for the Ricci scalar:
\begin{equation}
R = - 2 R_{lk} = - 2 R_{lklk}.
\end{equation}

The six amplitudes $\{\Psi_2,\Psi_3,\Psi_4,\Phi_{22}\}$ of a wave
are generally observer-dependent \cite{Eardley1973} (for more
details see the Apendix \ref{apendice}). However, there are
certain invariant statements about them that are true for all
standard observers if they are true for any one. These statements
characterize invariant $E(2)$ classes of waves. The name of each
class is composed of the Petrov type of its nonvanishing Weyl
tensor and the maximum number of nonvanishing amplitudes
$\{\Psi_2,\Psi_3,\Psi_4,\Phi_{22}\}$ (the dimension of
representation) as seen by any observer. Both the Petrov type and
the dimension of representation are independent of observer.
Considering standard observers such that: (a) each observer sees
the wave travelling in the $+z$ direction, and (b) each observer
measures the same frequency for a monochromatic wave, then the
$E(2)$ classes in order of decreasing generality are:
\begin{itemize}
 \item \textbf{Class} $II_6$: $\Psi_2 \neq 0$. All standard
     observers measure the same non-zero amplitude in the
     $\Psi_2$ mode. But the presence or absence of all other
     modes is observer-dependent;
 \item \textbf{Class} $III_5$: $\Psi_2 = 0,~\Psi_3 \neq 0$.
     All standard observers measure the absence of $\Psi_2$
     and the presence of $\Psi_3$. But the presence or absence
     of $\Psi_4$ and $\Phi_{22}$ is observer-dependent;
 \item \textbf{Class} $N_3$: $\Psi_2 = \Psi_3 = 0,~\Psi_4 \neq
     0, \Phi_{22} \neq 0$. Presence or absence of all modes is
     observer-independent;
 \item \textbf{Class} $N_2$: $\Psi_2 = \Psi_3 = \Phi_{22} =
     0;~\Psi_4 \neq 0$. Observer-independent;
 \item \textbf{Class} $O_1$: $\Psi_2 = \Psi_3 = \Psi_4 =
     0;~\Phi_{22} \neq 0$. Observer-independent;
 \item \textbf{Class} $O_0$: $\Psi_2 = \Psi_3 = \Psi_4 =
     \Phi_{22} = 0$. Observer-independent. All standard
     observers measure no wave.
\end{itemize}

\section{Polarization modes of gravitational waves in $f(R)$ theories}
\subsection{The metric formalism}
Let the gravitational action be an arbitrary function of the Ricci
scalar:
\begin{equation}\label{action f(R)}
I = \int d^4 x \sqrt{-g} f(R).
\end{equation}
By varying this action with respect to the metric $g_{\mu\nu}$ we
have the following  vacuum field equations:
\begin{equation}\label{field eq}
f^{\prime} R_{\mu\nu} - \frac{1}{2}f g_{\mu\nu} -
\nabla_\mu\nabla_\nu f^{\prime} + g_{\mu\nu} \Box f^{\prime}  =
0,
\end{equation}
where the prime represents derivatives with respect to $R$.

In the sequence we restrict ourselves to the following class of
the $f(R)$ theories:
\begin{equation}\label{f(R) class}
f(R) = R - \alpha R^{-\beta}
\end{equation}
Substituting (\ref{f(R) class}) in the field equations (\ref{field
eq}) we obtain the following relations between $R_{\mu\nu}$ and
the Ricci scalar $R$:
\begin{equation}\label{ricci eq}
R_{\mu\nu} = \frac{(R-\alpha R^{-\beta})g_{\mu\nu} + 2\alpha \beta \nabla_\mu \nabla_\nu R^{-(1+\beta)}
- 2\alpha\beta g_{\mu\nu}\Box R^{-(1+\beta)}}{2[1+\alpha\beta R^{-(1+\beta)}]}.
\end{equation}
Contracting this expression we have a dynamical equation for $R$:
\begin{equation}\label{escalar eq}
R = \alpha \left[(\beta + 2)R^{-\beta} + 3\beta\Box R^{-(1+\beta)}\right]
\end{equation}
The classification procedure involves examining the far-field,
linearized, vacuum field equations of a theory. In what follows we
examine different cases for the possible values of $\alpha$ and
$\beta$. In each case, we first find $R$ from equation
(\ref{escalar eq}) and then we can compute $R_{\mu\nu}$ from
(\ref{ricci eq}).
\\
\begin{itemize}
\item \emph{\textbf{Case}} $\alpha = 0$
\end{itemize}

This is the trivial case, which reduces to the general relativity
theory. From the equations (\ref{escalar eq}) and (\ref{ricci eq})
we find $R = 0$ and $R_{\mu\nu} = 0$. Consequently, from the
relations of the section \ref{revisit} we deduce that:
\begin{equation}
R_{lklk} = R_{lml\overline{m}} = R_{lklm} = R_{lkl\overline{m}} = 0
\end{equation}
and so we have:
\begin{equation}
\Psi_2 = \Psi_3 = \Phi_{22} = 0.
\end{equation}
And since we have no further constrains:
\begin{equation}
\Psi_4 \neq 0.
\end{equation}
And, as expected, the $E(2)$ classification for general relativity
is $N_2$.
\\
\begin{itemize}
\item \emph{\textbf{Case}} $\alpha\neq 0, \beta \geq 1$
\end{itemize}
For $\alpha \neq 0$, the equation (\ref{escalar eq}) can be
written as:
\begin{equation}\label{escalar eq 2}
\Box R^{-(1+\beta)} + \frac{\beta + 2}{3\beta} R^{-\beta} - \frac{1}{3\alpha \beta} R = 0.
\end{equation}
Working in the weak field regime, if $\beta \geq 1$ we have
$R^{-\beta} \gg R$ and the equation (\ref{escalar eq 2}) now
reads:
\begin{equation}\label{phi eq}
\Box \phi +\frac{\beta + 2}{3\beta} \phi^{\frac{\beta}{1+\beta}} = 0,
\end{equation}
where we have renamed $\phi \equiv R^{-(1+\beta)}$. But this
equation is of the form:
\begin{equation}\label{general phi eq}
\Box \phi - \frac{\partial U}{\partial \phi} = 0.
\end{equation}
Comparing (\ref{phi eq}) and (\ref{general phi eq}) we have the
potential:
\begin{equation}\label{potential case 2}
U(\phi) = -\left[\frac{(\beta+2)(\beta+1)}{3\beta(2\beta + 1)}\right]\phi^{\frac{2\beta + 1}{\beta + 1}}
\end{equation}
Since the field $\phi$ is Lorentz-invariant, we can solve the
equation (\ref{general phi eq}) by a very known method (see, e.g.,
\cite{Rajaramam1982}). First, consider the static solution of
(\ref{general phi eq}), i.e., the solution of the equation:
\begin{equation}
\frac{d^2 \phi}{d z^2} = \frac{\partial U}{\partial \phi},
\end{equation}
which can be written as:
\begin{equation}\label{virial}
\frac{1}{2}\left(\frac{d\phi}{dz}\right)^2 = U(\phi).
\end{equation}
Assuming the potential (\ref{potential case 2}) in (\ref{virial})
and integrating we have:
\begin{equation}\label{static solution 1}
\phi(z) = \left[ i \xi (z-z_0)+\phi_0^{1/2(\beta+1)}\right]^{2(\beta + 1)},
\end{equation}
where
\begin{equation}
\xi=\frac{1}{2(\beta + 1)}\left[\frac{2(\beta+2)(\beta+1)}
{3\beta(2\beta + 1)}\right]^{1/2}
\end{equation}
and $\phi_0 = \phi(z_0)$ is the value of the field $\phi$ in some
initial position $z_0$. Now, because the system is
Lorentz-invariant, given the static solution (\ref{static solution
1}), one can Lorentz-transform it to obtain the time dependent
solution:
\begin{equation}
\phi(z,t) = \left[ i \xi \frac{(z-z_0) - vt}{\sqrt{1-v^2}} + \phi_0^{1/2(\beta+1)} \right]^{2(\beta+1)}.
\end{equation}
And, therefore, the Ricci scalar reads:
\begin{equation}\label{escalar solution 1}
R(z,t) = \left[  i \xi \frac{(z-z_0) - vt}{\sqrt{1-v^2}} + R_0^{-1/2} \right]^{-2},
\end{equation}
where $v$ is the wave propagation velocity. This is the solution
of the equation (\ref{phi eq}) as can be verified by direct
substitution.

Substituting the solution (\ref{escalar solution 1}) in
(\ref{ricci eq}) we find the non-zero components of the Ricci
tensor to first order in $R$:
\begin{equation}
R_{tt} = \frac{1}{6\beta}\left[ (1-2\beta) - \frac{2(\beta + 2)v^2}{1-v^2}\right]R,
\end{equation}
\begin{equation}
R_{tz} = \frac{(\beta + 2)v}{3\beta(1-v^2)}R,
\end{equation}
\begin{equation}
R_{zz} = \frac{1}{6\beta}\left[(2\beta - 1) - \frac{2(\beta+2)}{1-v^2} \right] R.
\end{equation}
Therefore, from the relations of the section \ref{revisit} we
find:
\begin{equation}
R_{lklm} = R_{lkl\overline{m}} = 0,~~R_{lklk} \neq 0,~~R_{lml\overline{m}} \neq 0,
\end{equation}
and so:
\begin{equation}
\Psi_2 \neq 0, ~~\Psi_3 = 0, ~~\Psi_4 \neq 0 ~{\rm{and}}~ \Phi_{22} \neq 0,
\end{equation}
and the $E(2)$ classification for this case is $II_6$.
\\
\begin{itemize}
\item \emph{\textbf{Case}} $\alpha\neq 0, \beta < -2$
\end{itemize}
Considering now $\beta < -2$ we have $R^{-\beta} \ll R$ and the
equation (\ref{escalar eq 2}) reads:
\begin{equation}
\Box \phi - \frac{1}{3\alpha \beta} \phi^{-1/(1+\beta)} = 0,
\end{equation}
where $\phi$ has the same definition presented above. However, now
the potential is given by:
\begin{equation}
U(\phi) = \frac{1}{3\alpha\beta}\phi^{\frac{\beta}{\beta + 1}},
\end{equation}
from which we obtain the static solution:
\begin{equation}
\phi(z) = \left[\zeta (z-z_0) + \phi_0^{\frac{\beta+2}{2(\beta+1)}} \right]^{\frac{2(\beta+1)}{\beta+2}},
\end{equation}
where:
\begin{equation}
\zeta = \frac{\beta+2}{(\beta+1)\sqrt{6\alpha\beta}}.
\end{equation}
And after a Lorentz transformation we find the full solution:
\begin{equation}
\phi(z,t) = \left[ \zeta \frac{(z-z_0) - vt}{\sqrt{1-v^2}} + \phi_0^{\frac{\beta+2}{2(\beta+1)}} \right]^{\frac{2(\beta+1)}{\beta+2}}.
\end{equation}
Thus, the evolution of the Ricci scalar for this case is:
\begin{equation}
R(z,t) = \left[\zeta\frac{(z-z_0) - vt}{\sqrt{1-v^2}} + R_0^{-\frac{\beta+2}{2}} \right]^{-\frac{2}{\beta+2}}.
\end{equation}
The components of the Ricci tensor reads:
\begin{equation}
R_{tt} = \frac{1}{3}\left[\frac{\beta}{(\beta+1)}\frac{v^2}{(1-v^2)} - \frac{1}{2} \right]R,
\end{equation}
\begin{equation}
R_{tz} = -\frac{\beta}{3(\beta + 1)}\frac{v}{(1-v^2)}R,
\end{equation}
\begin{equation}
R_{zz} = \frac{1}{3} \left[\frac{\beta}{(\beta+1)}\frac{1}{(1-v^2)} + \frac{1}{2}\right]R,
\end{equation}
while all the others components are null.
\\
Therefore, together  with the formalism presented in the section
\ref{revisit}, we can deduce that:
\begin{equation}
\Psi_2 \neq 0, ~~\Psi_3 = 0, ~~\Psi_4 \neq 0 ~{\rm{and}}~ \Phi_{22} \neq 0.
\end{equation}
And we are lead to classify the theories with $\beta < -2$ in the
$E(2)$ class $II_6$.
\\
\begin{itemize}
\item \emph{\textbf{Case}} $\alpha\neq 0, \beta = -2$
\end{itemize}
This is a particular case, where the behaviour of the Ricci scalar
and Ricci tensor are oscillatory. That is, if we use $\beta=-2$ in
the equation (\ref{escalar eq 2}), we obtain:
\begin{equation}
\Box R -\frac{1}{6\alpha} R = 0,
\end{equation}
with the solution:
\begin{equation}
R = R_0 \exp(i k_\alpha x^\alpha), ~~~~k_\alpha k^\alpha = \frac{1}{6\alpha}.
\end{equation}
Considering this solution in the equation (\ref{ricci eq}) with
$R\ll 1$ we find the non-null components of the Ricci tensor:
\begin{equation}
R_{tt} = \frac{1}{2}(4\alpha k^2 - 1)R
\end{equation}
\begin{equation}
R_{tz} = -2\alpha k \sqrt{k^2 - \frac{1}{6\alpha}}R
\end{equation}
\begin{equation}
R_{zz} = \frac{1}{6}(12\alpha k^2 + 1)R
\end{equation}
So, again we have:
\begin{equation}
\Psi_2 \neq 0, ~~\Psi_3 = 0, ~~\Psi_4 \neq 0 ~{\rm{and}}~ \Phi_{22} \neq 0,
\end{equation}
and the $E(2)$ classification is $II_6$.
\\
As can be seen, for all the studied cases (except the case $\alpha
= 0$ ), the theory given by eq.(\ref{f(R) class}) is classified in
the class $II_6$, i.e., the most general classification where all
the six polarization can appear for some specific Lorentz
observer, but the amplitude $\Psi_2$ is observer-independent.

\subsection{The Palatini approach}
In the Palatini approach, the metric $g$ and the (usually
torsionless) connection $\Gamma$ are considered as independent
variables, entering the definition of the Ricci tensor. The vacuum
field equations, derived from the Palatini variational principle
applied in the action (\ref{action f(R)}) are:
\begin{equation}\label{palatini field}
f^{\prime}R_{\left(\mu\nu \right)} - \frac{1}{2}fg_{\mu\nu} = 0,
\end{equation}
\begin{equation}
\nabla_\alpha^{\Gamma}(\sqrt{-g}f^\prime g^{\mu\nu}) = 0,
\end{equation}
where $\nabla_\alpha^{\Gamma}$ is the covariant derivative with
respect to $\Gamma$. We shall use the standard notation denoting
by $R_{(\mu\nu)}$ as the symmetric part of $R_{\mu\nu}$, i.e.
$R_{\left(\mu\nu\right)}\equiv \frac{1}{2} \left(R_{\mu\nu} +
R_{\nu\mu}\right)$.
It was shown that the vacuum field equations
(\ref{palatini field}) leads to `universal' equations for a wide
range of functions $f(R)$ \cite{ferraris1994}. These universal
equations are just Einstein equations with cosmological constant
$\Lambda$.
\\
Thus, the properties of vacuum GWs in the $f(R)$ gravity using the
Palatini approach reduces to the problem of GWs in the Einstein
equations considering the cosmological term. In a recent work,
N\"af \emph{et.al} \cite{naf2009} have analyzed this case. They
expanded the perturbations in a de Sitter and an anti-de Sitter
background. Since the Minkowski metric is not a solution of the
vacuum field equations this approach seems to be the most
straightforward. Considering terms up to linear order in $\Lambda$
they calculated the non-null components of the Riemann tensor and
found that $\Lambda$ does not introduce additional polarization
states for the GWs. Moreover, they shown that the cosmological
term introduces tiny modifications in the amplitude of the wave
which are well below the detectability of the present GWs
detectors.
\\
Therefore, we can conclude that GWs in the $f(R)$ gravity using
the Palatini approach has only the two usual polarizations of
general relativity, i.e., the polatization $+$ and $\times$.

\section{Polarization modes of gravitational waves in quadratic gravity}

For completeness we analyze the polarizations for the following
Lagrangian of quadratic gravity:
\begin{equation}
I = \int d^4x\sqrt{-g} \left[ R + \alpha R^2 + \gamma R_{\mu\nu}R^{\mu\nu} \right].
\end{equation}
Variation of this action in respect to the metric gives the field
equations:
\begin{equation}\label{quadratic field}
G_{\mu\nu} + \alpha H_{\mu\nu} + \gamma I_{\mu\nu} = 0,
\end{equation}
where:
\begin{equation}
H_{\mu\nu} = 2 \nabla_\mu\nabla_\nu R - 2g_{\mu\nu}\Box R +\frac{1}{2}g_{\mu\nu}R^2 - 2RR_{\mu\nu}
\end{equation}
and
\begin{equation}
I_{\mu\nu} = \nabla_\mu\nabla_\nu R -\frac{1}{2}g_{\mu\nu} \Box R - \Box R_{\mu\nu}
- 2{R_\mu}^\alpha R_{\alpha\nu} + \frac{1}{2}g_{\mu\nu}R_{\alpha\beta}R^{\alpha\beta}
\end{equation}
The field equations (\ref{quadratic field}) were analyzed in the
linearized regime by de Rey Neto \emph{et. al}
\cite{deReyNeto2003} using a perturbative approach developed from
the concept of regular reduction of a system of partial
differential equations. Working only with the transverse-traceless
part of the metric perturbations $h_{ij}^{TT}$, they found a
frequency-dependent correction in the gravitational wave amplitude
due to the presence of the Ricci-squared term in the gravitational
action.
\\
Here, we do not write explicitly the field equations in terms of
the perturbations. Instead, we consider the Ricci tensor and the
Ricci scalar as first order functions of the general metric
perturbations $h_{\mu\nu}$. Then, we find solutions for the
dynamical equations to linear order in $R_{\mu\nu}$ and $R$. These
solutions enable us to find the non-null Newman-Penrose quantities
and classify the quadratic gravity in analogy with was made for
$f(R)$ theories in the preceding section.
\\
Considering the equation (\ref{quadratic field}) to linear order
in $R$ and $R_{\mu\nu}$ we have the following equation for the
Ricci tensor:
\begin{equation}\label{quadratic ricci}
\Box R_{\mu\nu} -\frac{1}{\gamma}R_{\mu\nu} = \frac{1}{\gamma} S_{\mu\nu},
\end{equation}
where:
\begin{equation}
S_{\mu\nu} = (2\alpha + \gamma)\left[\partial_\mu\partial_\nu R - \frac{\eta_{\mu\nu}R}{4(3\alpha + \gamma)}\right]
\end{equation}
Taking the trace of (\ref{quadratic ricci}) we find:
\begin{equation}\label{quadratic scalar}
2(3\alpha + \gamma)  \Box R + R = 0.
\end{equation}
From this equation, we can find that a particular combination of
parameters, namely $\gamma = -3\alpha$, leads to $R = 0$. For this
case the solution of equation (\ref{quadratic ricci}) reads:
\begin{equation}
R_{\mu\nu} = A_{\mu\nu} \exp(iq_\alpha x^\alpha),~~~~q_\alpha q^\alpha = -\frac{1}{\gamma}.
\end{equation}
Since there are no further constrains we find:
\begin{equation}
R_{lklk} = 0,~~R_{lml\overline{m}} \neq 0,~R_{lklm} \neq 0,~R_{lkl\overline{m}} \neq 0.
\end{equation}
Therefore:
\begin{equation}
\Psi_2 = 0, ~~\Psi_3 \neq 0, ~~\Psi_4 \neq 0 ~{\rm{and}}~ \Phi_{22} \neq 0,
\end{equation}
and the correspond $E(2)$ class is $III_5$.
\\
For the case $\gamma \neq -3\alpha$, the equation (\ref{quadratic
scalar}) gives:
\begin{equation}
R = R_0 \exp(ik_\alpha^1 x^\alpha),~~k^1_\alpha k_1^\alpha = \frac{1}{2(3\alpha + \gamma)}.
\end{equation}
With this expression in (\ref{quadratic ricci}), the full solution
for the Ricci tensor reads:
\begin{equation}\label{quadratic ricci answer}
R_{\mu\nu} = A_{\mu\nu} e^{i(k_1z - \omega t)} + B_{\mu\nu}  e^{i(k_2 z - \omega t)} + c.c.,
\end{equation}
where:
\begin{equation}
A^{\mu\nu} = \frac{2}{3}R_0(3\alpha + \gamma) \left[k_1^\mu k_1^\nu + \frac{\eta^{\mu\nu}}{4(3\alpha + \gamma)}\right],
\end{equation}
and:
\begin{equation}
k_1 = \sqrt{\omega^2 + \frac{1}{2(3\alpha + \gamma)}},~~k_2 = \sqrt{\omega^2 -\frac{1}{\gamma}}.
\end{equation}
Now, from (\ref{quadratic ricci answer}) we write explicitly all
the components of $R_{\mu\nu}$:
\begin{equation}
R_{tt} = \frac{1}{6}\left[4(3\alpha +\gamma)\omega^2 - 1\right]R_0 e^{i(k_1 z - \omega t )} + B_{tt} e^{i(k_2 z -\omega t)} + c.c.,
\end{equation}
\begin{equation}
R_{tz} = -\frac{2}{3}(3\alpha + \gamma) \omega \sqrt{\omega^2 + \frac{1}{2(\gamma + 3\alpha)}} R_0 e^{i(k_1 z - \omega t )} + B_{tz} e^{i(k_2 z -\omega t)} +c.c.,
\end{equation}
\begin{equation}
R_{zz} = \frac{1}{2}\left[1 + \frac{4}{3}(3\alpha + \gamma)\omega^2\right]R_0 e^{i(k_1 z - \omega t )} + B_{zz}e^{i(k_2 z -\omega t)} + c.c.,
\end{equation}
and all the other components satisfies:
\begin{equation}
R_{ij} = B_{ij} e^{i(k_2 z -\omega t)} + c.c.,
\end{equation}
where $i,j = x,y,z$.
\\
Therefore, since there are no further constrains on $B_{\mu\nu}$,
all the components of the Riemann tensor in the tetrad basis are
non-null:
\begin{equation}
R_{lklk} \neq 0,~R_{lml\overline{m}} \neq 0,~R_{lklm} \neq 0,~R_{lkl\overline{m}} \neq 0,
\end{equation}
and so, all the Newman-Penrose quantities are also non-null:
\begin{equation}
\Psi_2 \neq 0, ~\Psi_3 \neq 0, ~~\Psi_4 \neq 0 ~{\rm{and}}~ \Phi_{22} \neq 0.
\end{equation}
Thus, the $E(2)$ classification for the quadratic gravity in the
most general case is $II_6$.

\section{Conclusions}

Recently, working in the metric formalism, Cappoziello et al.
\cite{Capozziello} showed that GWs in $f(R)$ gravity can have a
massive-like scalar mode and a longitudinal force besides the two
polarizations which appear in general relativity. This agrees with
our result that the quantities $\Phi_{22}$ and $\Psi_2$ are
non-null for the particular class $f(R) = R + \alpha R^{-\beta}$.
The $\Phi_{22}$ and $\Psi_2$ amplitudes correspond, respectively,
to a perpendicular scalar mode (breathing mode) and to a
longitudinal scalar mode. However, it is worth emphasizing that
since the $\Psi_2$ mode are non-null, the $E(2)$ classification is
$II_6$ (see section \ref{revisit}). So, for this class, it is
always possible to find a Lorentz observer who measures all the
six polarization states. On the other hand, GWs in the Palatini
approach have only the two usual polarizations states such as
general relativity.
\\
Furthermore, the method we use is not only simple but very robust
and we are able to obtain some important information regarding the
theories considered here. The key observational GW amplitude is
$\Psi_2$. If the amplitude $\Psi_2$ would be detected, the $f(R)$
models in the metric formalism like the one considered here would
be supported and so the quadratic gravity for $\gamma \neq
-3\alpha$. In this case we need another method to distinguish the
two theories, we could compare the wave form in both cases, for
example. If the $\Psi_3$ mode would be detected, but not the
amplitude $\Psi_2$, only the quadratic gravity for $\gamma =
-3\alpha$ would be supported.
%Otherwise, if only the $\Psi_4$ mode would be detected, general
%relativity theory would be supported and also the Palatini
%approach for $f(R)$ theories.
\\
Therefore, if we would be able to detect GWs, an important way to
identify the theory of gravity could be established
\cite{corda2009}. In the particular case of $f(R)$ gravity in the
Palatini approach, we showed that the polarizations of GWs are the
same of general relativity. However, it is worth stressing that
other information contained in the GW signals, like waveform and
phase of the signal, could be important to permit, together with
the polarizations, a clear identification of the theory.

\section*{Acknowledgments}
The authors thank Prof. F\'abio C. Carvalho and Ant\^onio C.A.
Faria for helpful discussions and for clarifying some important
points. MESA would like to thank the Brazilian Agency FAPESP for
support (grant 06/03158-0). ODM and JCNA would like to thank the
Brazilian agency CNPq for partial support (grants 305456/2006-7
and 307424/2007-3 respectively).

\appendix
\section{Behavior of the NP amplitudes under Lorentz transformations}\label{apendice}

We can understand the $E(2)$ classification scheme by analyzing
the behavior of the Newman-Penrose (NP) amplitudes
$\{\Psi_2,\Psi_3,\Psi_4,\Phi_{22}\}$ under a Lorentz
transformation of the complex tetrad basis. This point is well
explained in the reference \cite{Eardley1973}. Here we summarize
the main idea.
\\
Consider two standard observers $O$ and $O^\prime$ with tetrads
$(\textbf{k},\textbf{l},\textbf{m},\textbf{n})$ and
$(\textbf{k}^\prime,\textbf{l}^\prime,\textbf{m}^\prime,\textbf{n}^\prime)$
respectively. If we choose the $\textbf{k}$ as proportional to the
wave vector (following the convention of Eardley et al.), so we
have $\textbf{k} = \textbf{k}^\prime$. The most general proper
Lorentz transformation relating the tetrads that keep $\textbf{k}$
fixed is:
\begin{equation}
\textbf{k}^\prime = \textbf{k},
\end{equation}
\begin{equation}
\textbf{m}^\prime = e^{i\varphi}(\textbf{m} + \sigma\textbf{k}),
\end{equation}
\begin{equation}
{\overline{\textbf{m}}}^\prime = e^{-i\varphi}(\overline{\textbf{m}} + \overline{\sigma}\textbf{k}),
\end{equation}
\begin{equation}
{\textbf{l}}^\prime = {\textbf{l}} + \overline{\sigma}{\textbf{m}} + \sigma \overline{\textbf{m}} + \sigma \overline{\sigma} {\textbf{k}},
\end{equation}
where $\sigma$ is an arbitrary complex number which produces null
rotations (particular combinations of boosts and rotations), while
$\varphi$, which runs from $0$ to $2\pi$, is an arbitrary real
phase that produces a rotation about $\textbf{e}_z$.
\\
The transformations induced on the amplitudes of a wave by
($\varphi$, $\alpha$) is:
\begin{equation}\label{psi 2}
\Psi_2^\prime = \Psi_2,
\end{equation}
\begin{equation}
\Psi_3^\prime = e^{-i\varphi}(\Psi_3 + 3\overline{\sigma}\Psi_2),
\end{equation}
\begin{equation}
\Psi_4^\prime = e^{-2i\varphi}(\Psi_4 + 4\overline{\sigma}\Psi_3 + 6 \overline{\sigma}^2 \Psi_2),
\end{equation}
\begin{equation}\label{phi 22}
\Phi_{22}^\prime = \Phi_{22} + 2\sigma\Psi_3 + 2 \overline{\sigma}\overline{\Psi}_3 + 6\sigma\overline{\sigma}\Psi_2.
\end{equation}
Now, it is evident from this set of equations that the amplitudes
$\{\Psi_2,\Psi_3,\Psi_4,\Phi_{22}\}$ cannot be specified in an
observer-independent manner. For example, suppose that the
observer in $O$ measure a wave having the only nonvanishing
amplitude $\Psi_2$ ($s = 0$). The observer in $O^\prime$, in
relative motion with respect to $O$, will conclude that the wave
has the nonvanishing amplitudes $\Psi_2$, $\Psi_3$, $\Psi_4$ and
$\Phi_{22}$ ($s = 0, \pm 1, \pm 2, 0$). However, there is a set of
invariant statements which define the $E(2)$ classification
scheme. Thus, we classify waves in an $E(2)$ invariant manner by
uncovering all representations of $E(2)$ embodied in equations
(\ref{psi 2})$-$(\ref{phi 22}). Each such representation, in which
some of the NP amplitudes vanish identically, is a distinct
invariant class. The description of each class can be found in the
main text.


\begin{thebibliography}{10}

\bibitem{Weyl1922} H. Weyl, Space-Time-Matter, 4th edn Dover, New York
(1922).
\bibitem{deWitt1965} B. S. De Witt, The Dynamical Theory of
Groups and Fields, Gordon and Breach, New York (1965).
\bibitem{Hu2004} B. L. Hu, E. Verdaguer, Living Rev.
Relativity 7 (2004) 3.
\bibitem{Birrel1982} N. D. Birrel, P. C. W. Davies, Quantum Fields in Curved Space, Cambridge University Press,
Cambridge (1982).
\bibitem{stelle1977} K. S. Stelle, Phys. Rev. D 16 (1977) 953.
\bibitem{Zwiebach1985} B. Zwiebach, Phys. Lett. B 156 (1985) 315.
\bibitem{Tseytlin1986} A. Tseytlin, Phys. Lett. B 176 (1986) 92.
\bibitem{Starobinsky1980} A. A. Starobinsky, Phys. Lett. B
91 (1980) 99.
\bibitem{several} S. Capozziello, M. Francaviglia, Gen. Relativ.
Gravit. 40 (2008) 357.
%\bibitem{several} Nojiri, S., and Odintsov, S. D. 2003, Phys. Rev.
%D 68, 123512; 2003, Phys. Lett. B 576, 5; 2004, Gen. Relativ.
%Gravit. 36, 1765; Carroll, S. M., Duvvuri, V., Trodden, M., and
%Turner, M. S. 2004, Phys. Rev. D 70, 043528; Meng, X. H., and
%Wang, P. 2004, Phys. Lett. B 584, 1; Capozziello, S., Cardone, V.
%F. and Troisi, A. 2005, Phys. Rev. D 71, 043503;  Cruz-Dombriz A.,
%and Dobado, A. 2006, Phys. Rev. D 74, 087501; Nojiri, S., and
%Odinstsov, S. D. 2007, Int. J. Geom. Methods Mod. Phys. 4, 115;
%Capozziello, S., and Francaviglia, M., arXiv:0706.1146v1; Appleby,
%S. A., and Battye, R., arXiv:0705.3199; Starobinsky, A. A.,
%arXiv:0707.2041
\bibitem{deReyNeto2003}E. C. de Rey Neto, O. D. Aguiar, J. C. N. de
Araujo, Class. Quantum Grav. 20 (2003) 2025.
\bibitem{Capozziello}C. Corda, JCAP 0704 (2007) 009; Int. J.
Mod. Phys. A 23 (10) (2008), 1521; S. Capozziello, C. Corda, M. F.
De Lautentis, Phys. Lett. B 669 (2008) 255.
\bibitem{Eardley1973}D. M. Eardley, D. L Lee, A. P. Lightman, Phys. Rev. D 8 (1973)
3308.
\bibitem{Newman}E. Newman, R. Penrose, J. Math Phys. 3 (1962) 566; see errata, \emph{ibid.} 4 (1962)
998.
\bibitem{virgo2007}F. Acernese, et al., Virgo Collaboration,
Class. Quantum Grav. 24 (19) (2007) S381.
\bibitem{ligo}D. Sigg, for the LIGO Collaboration, http://www.ligo.org/pdf\_public/P050036.pdf
\bibitem{ligo2005}B. Abbott, et al., LIGO Collaboration, Phys.
Rev. D 72 (2005) 042002
\bibitem{tama2002}M. Ando, TAMA Collaboration, Class. Quantum
Grav. 21 (5) (2002) 1615
\bibitem{aguiar2008}O. D. Aguiar, et al., Class. Quantum Grav. 25
(11) (2008) 114042.
\bibitem{dePaula2005}W.L.S. {de Paula}, O.D. Miranda, R.M.
Marinho, Class. Quantum Grav. 21 (2004) 4595
\bibitem{Will2006} C. M. Will, Living Reviews in Relativity (2006)
http://relativity.livingreviews.org/Articles/Irr-2006-3
\bibitem{Rajaramam1982} R. Rajaraman, Solitons and
Instantons: An Introduction to Solitons and Instantons in Quantum
Field Theory, Elsevier Science Publishers, Netherlands (1982).
\bibitem{ferraris1994}M. Ferraris, M. Francaviglia, I. Volovich, Class. Quantum Grav. 11 (1994) 1505
\bibitem{naf2009}J. N\"af, P. Jetzer, M. Sereno, Phys. Rev.
D 79 (2009) 024014
\bibitem{corda2009} C. Corda, 2009, arXiv: 0905.2502
%\bibitem{Albrecht2000} Albrecht, A., and Skordis, C. 2000, \PRL, 2076, 84




\end{thebibliography}
\end{document}